\documentclass[showpacs,twocolumn,prl,superscriptaddress]{revtex4-1}

\usepackage{slashed}
\usepackage{mathrsfs}

\usepackage{amsmath}
\usepackage{amssymb}
\usepackage{revsymb}
\usepackage{graphicx,accents}
\usepackage{graphicx,epstopdf}
\usepackage{mathrsfs}
\usepackage{bm}
\usepackage{psfrag}
\usepackage[usenames,dvipsnames]{xcolor}
\usepackage{hyperref}
\hypersetup{colorlinks=true,citecolor=Red,urlcolor=Red}
\usepackage{bbm}
\usepackage{bbold}
\usepackage{tabularx}
\usepackage{graphicx}
\usepackage[normalem]{ulem}

\usepackage{multirow,tabularx}
\usepackage{xcolor,pifont}

\newcommand*\colourcheck[1]{%
  \expandafter\newcommand\csname #1check\endcsname{\textcolor{#1}{\ding{52}}}%
}

\newcommand{\be}{\begin{equation}}
\newcommand{\ee}{\end{equation}}
\newcommand{\bea}{\begin{eqnarray}}
\newcommand{\eea}{\end{eqnarray}}

\newcommand{\eps}{\varepsilon}
\newcommand{\mbf}[1]{\mathbf{#1}}
\newcommand{\trm}[1]{\textrm{#1}}

\newcommand{\tsf}[1]{\textsf{#1}}

\newcommand{\vkap}{\varkappa}
\newcommand{\vphi}{\varphi}

\newcommand{\e}{\mbox{e}\,}

\newcommand{\LCperp}{{\scriptscriptstyle \perp}}

\begin{document}


\title{Highly polarised gamma photons from electron-laser collisions}

\author{S.~Tang}
\email{suo.tang@plymouth.ac.uk}
\affiliation{Centre for Mathematical Sciences, University of Plymouth, Plymouth, PL4 8AA, United
Kingdom}
\author{B.~King}
\email{b.king@plymouth.ac.uk}
\affiliation{Centre for Mathematical Sciences, University of Plymouth, Plymouth, PL4 8AA, United
Kingdom}
\author{H.~Hu}
\affiliation{Hypervelocity Aerodynamics Institute, China Aerodynamics Research and Development Center, 621000 Mianyang, Sichuan, China}


\date{\today}
\begin{abstract}
We detail a method to produce a GeV-photon source with polarisation purity exceeding $96\%$ and $89\%$ for linear and circular polarisation respectively and with a brilliance of the order of $10^{21}\,\textrm{photons}/(\textrm{s}~\textrm{mm}^2~\textrm{mrad}^2~0.1\%~\textrm{BW})$. Using currently available multi-GeV electron bunches and laser pulses of moderately relativistic intensities, we show how the weakly nonlinear regime can produce photons polarised mainly parallel to the laser field. We demonstrate the robustness of this scheme by considering electron bunches of various emissivities colliding with linearly and circularly-polarised laser pulses at a range of angles.
\end{abstract}
\maketitle

When an accelerated electron bunch scatters off a laser pulse of sufficient intensity, photons are produced in a series of harmonics corresponding to the number of absorbed laser photons. As the intensity of the pulse is increased, there is predicted to be a transition from the perturbative \emph{multi-photon} regime, where spectra are well-approximated by considering just the lowest harmonics, to the \emph{small-coupling non-perturbative} regime where all orders of harmonics can contribute equally. Experiments such as {LUXE} at DESY \cite{Abramowicz:2019gvx} and E320 at FACET-II \cite{slacref1} are planned in the near future to combine electron bunches of order $10\,\trm{GeV}$ with laser pulses of intensity parameter $\xi\sim O(1)$ to measure,
for the first time, the transition from the multi-photon to the non-perturbative regime of quantum electrodynamics (QED). They will thereby complement the landmark E144 experiment, which measured processes in the multi-photon regime two decades ago \cite{bula96,burke97,bamber99}. The measurement and characterisation of the non-perturbative regime is highly relevant to the design and analysis of experimental campaigns planned at the next generation of high power laser facilities such as ELI-Beamlines, ELI-NP \cite{eli16}, EP-OPAL and SEL \cite{shen18}, which will realise this regime in intense laser-plasma interactions.

Photon polarisation is an important experimental observable, which was central to measurements providing the first evidence of real photon-photon scattering (vacuum birefringence) from an isolated neutron star~\cite{mignani17}, and measuring the polarisation of X-ray sources is a key part of the future Imaging X-ray Polarimetry Explorer (IXPE)~\cite{ixpe16}. Having access to a highly-polarised source of photons has also been shown to significantly reduce the experimental demands required to provide the first experimental measurement of real photon-photon scattering using lasers in the lab \cite{king16,ilderton16,meuren17b}. The decay of a photon into an electron-positron pair in an intense electromagnetic field, which is believed to play an important role in the evolution of the magnetospheres of some neutron stars \cite{harding06}, and a key observable in the LUXE and E320 experiments, can also be considerably enhanced by using a highly-polarised photon source \cite{king13a}. Using pair spectroscopy \cite{homma17}, the polarisation purity can form a useful further interrogation of high-intensity QED. Furthermore, polarised photons also find application in e.g. the study of nuclear structure via photonuclear reactions~\cite{horikawa2014neutron}.

In this letter, we propose a robust scheme to generate brilliant and highly-polarised GeV $\gamma-$rays collimated by colliding an electron bunch nearly head-on with a laser pulse of intermediate intensity.
By considering an electron bunch with a divergence of $0.2\,$mrad \cite{PRL2019Petawatt}, the photon source is predicted to have a polarisation purity of around $96\%$ and $89\%$ for linear and circular polarisation respectively, and a brilliance of up to
$10^{21}~\textrm{photons}~\textrm{s}^{-1}~\textrm{mm}^{-2}~\textrm{mrad}^{-2}~0.1\%~\textrm{BW}$. Because our scheme exploits the harmonic and angular structure of the spectrum, it is beyond analyses based on numerical simulations that  employ the standard locally-constant-field approximation \cite{ritus85,king15d,dipiazza18,king19a}. For this reason, the predicted brilliance of our scheme is orders of magnitude higher than hitherto conceived for a polarised source~\cite{yanfei19}. Furthermore, as we will show, our scheme is robust and does not rely upon fine-tuning of experimental parameters.

We begin by outlining definitions used in the calculation. In an electron-laser collision, the differential probability of emitting a photon in the polarisation state $\eps_{j}$ with momentum $k$ via the nonlinear Compton process \cite{serbo04,king13a}, can be written as~\cite{seipt2017volkov}
\begin{align}
\frac{d^3\tsf{P}_{j}}{ds\,d\bm{r}^2} = \frac{\alpha}{(2\pi\eta)^{2}}\frac{s}{t}\int d\phi\,d\phi'~\tsf{T}_{j}\e^{i\int^{\phi}_{\phi'}\frac{k\cdot \pi_{p}(\vphi)}{m^2t\eta}d\vphi}\,,\label{eqn:sfi1}
\end{align}
where we model the laser pulse as a plane wave with wavevector $\vkap=\vkap^{0}(1,0,0,1)$,
$\alpha$ is the fine-structure constant, $\eta= \vkap\cdot p/m^{2}$, $s=\vkap\cdot k /\vkap\cdot p$ is the lightfront momentum fraction of the scattered photon, $t=1-s$, $\phi$ ($\phi'$) is the laser pulse phase,
$\pi_{p}=p-a+\vkap(2\,p\cdot a - a^{2})/\vkap\cdot p$, $a=eA$ where $A$ is the laser pulse vector potential, $p$, $e$ and $m$ are the electron incident momentum, absolute charge and mass respectively and $\tsf{T}_{j}$ is a polarisation-dependent integrand defined later. $\bm{r}=\mbf{k}^{\LCperp}/(sm)$ is the normalised photon momentum in the plane perpendicular to the laser propagation direction and relates directly to the scattering angle of the photon.
For later use, we define the angles to the negative $z$ axis, $\theta_{x,y}$, of the scattered photon in the $x$-$z$ and $y$-$z$ planes, via $r_{x,y}=m\eta\tan(\theta_{x,y}/2)/\vkap^0$.
The photon polarisation states are chosen to be the eigenstates of the polarisation operator in the given laser background to ensure the polarisation does not change after the photon is created. For a linearly-polarised laser pulse: $a(\phi)=m \xi \eps_{1} \sin(\phi) f(\phi)$ where $\xi$ and $f(\phi)$ are the laser amplitude and profile, and we use the normalised transverse states \cite{baier75a}:
\begin{align}
\epsilon_{1} = \eps_{1} - \frac{k\cdot \eps_{1}}{k\cdot \vkap} \vkap\,,~~~&\epsilon_{2} = \eps_{2}  - \frac{k\cdot \eps_{2} }{k\cdot \vkap}\vkap,\label{eqn:polbasis}
\end{align}
where $\eps_{1} = (0,1,0,0)$ and $\eps_2=(0,0,1,0)$ are parallel to the laser electric and magnetic fields respectively. For a circularly-polarised background $a(\phi)=m\xi [\eps_{1}\cos(\phi)+\eps_{2} \sin(\phi)]f(\phi)$, we use $\epsilon_{\pm} = (\epsilon_{1}\pm i\epsilon_{2})/\sqrt{2}$, where the sign $+$ ($-$) denotes the right-hand (left-hand) rotation of the polarisation. We will generally refer to $\epsilon_{1}$ and $\epsilon_{+}$ states ($\epsilon_{2}$ and $\epsilon_{-}$ states) as \emph{$E$-polarised} (\emph{$B$-polarised}) and as being parallel (perpendicular) to the field, even though, in general, the photon is not emitted head-on with the laser pulse, but instead at a small opening angle and so its polarisation direction is not exactly aligned with the field. We find:
\begin{subequations}
\begin{align}
\tsf{T}_{j} &= \frac{s^2}{8t} \Delta +w_{j}(\phi)\cdot w_{j}(\phi')\,,\\
\tsf{T}_{\pm} &=\frac{s^2}{8t} \Delta+\frac{1}{2}\bm{w}(\phi)\cdot\bm{w}(\phi')\pm if_t\bm{w}(\phi)\times \bm{w}(\phi')\,,
\end{align}
\label{Eq_trace}
\end{subequations}
$\!\!$where $\bm{w}(\phi) =(\bm{r}-\bm{p}^{\LCperp}/m)R(\phi) + \bm{a}^{\LCperp}(\phi)/m$, $\Delta=[a(\phi)-a(\phi')]^2/m^2$, $f_t=(1+t^2)/(4t)$ and $R(\phi) = 1-k\cdot\pi(\phi)/k\cdot p$ is the regulator (see e.g. \cite{king19d} for details).

\begin{figure}[t!!]
 \includegraphics[width=0.48\textwidth]{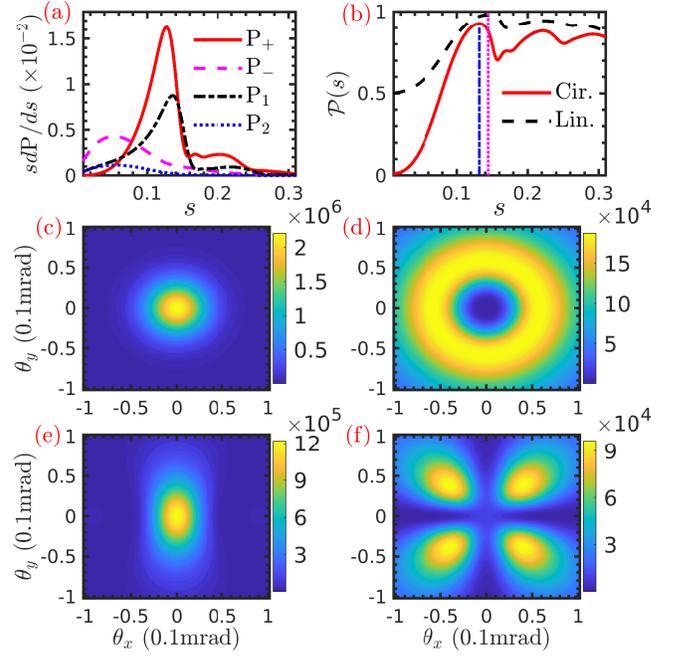}
\caption{Energy and angular spectra for a head-on collision of a $10\,\trm{GeV}$  ($\eta=0.095$) electron beam with a $\xi=0.5$, $\omega_l=1.24~\textrm{eV}$, $26.7\,\trm{fs}$ ($8$-cycle) laser pulse with envelope $f(\phi)=\cos^2(\phi/4\sigma)$ ($|\phi|\leq 2\pi\sigma$). Upper panels: (a) energy spectrum of the polarised photon $s\,d\tsf{P}_j/ds$ and (b) polarisation purity $\mathcal{P}(s)$. Central panels: angular distribution $d^2\tsf{P}/d\theta_xd\theta_y$ in a circularly-polarised background of $E$-polarised photons (c) and $B$-polarised photons (d). Bottom panels: angular distribution in a linearly-polarised background of $E$-polarised photons (e) and $B$-polarised photons (f). The blue (magenta) dot-dashed line in (b) denotes the Compton edge in a linearly (circularly) polarised background. }
\label{Fig_energy_Angle}
\end{figure}
In Eq.~(\ref{Eq_trace}), we see the polarisation-dependent part of the scattering probability for a linearly polarised background $\tsf{T}_{j}$ and for a circularly-polarised background $\tsf{T}_{\pm}$ is quite different. This dependence can lead to a significant difference in the energy and angular distribution of the scattered polarised photons.
We use the \emph{polarisation purity} $\mathcal{P}$, which is the fraction of photons in the $E$-polarised states, to quantify the differences: $\mathcal{P}=\tsf{P}_{1}/(\tsf{P}_{1}+\tsf{P}_{2})$ [$\mathcal{P}=\tsf{P}_{+}/(\tsf{P}_{+}+\tsf{P}_{-})$] for a linearly [circularly] polarised background. As an example in Fig.~\ref{Fig_energy_Angle}, we consider the head-on collision between a $10~\textrm{GeV}$  ($\eta=0.095$) electron and a $26.7\,\trm{fs}$ ($8$-cycle) laser pulse with intermediate intensity $\xi=0.5\,$.
Fig.~\ref{Fig_energy_Angle}(a) shows the different behaviour of the polarised energy spectra in the linearly and circularly polarised background for different values of $s$.
In the $s\to 0$ limit, the photons are almost unpolarised in a linearly-polarised background, whereas in a circularly-polarised background, almost all photons are $B$-polarised ($\mathcal{P}\to 0$).
Recalling that $s\propto 1+\cos\theta_{p}$, where $\theta_{p}\approx(\theta_{x}^{2}+\theta_{y}^{2})^{1/2}$ is the photon scattering angle with respect to the negative $z$-axis, we see that, although there is a source of highly polarised photons when $s\to 0$, this corresponds to: i) low energies and ii) broad angular spread.
However, another source of highly-polarised photons is at higher $s$-values starting around the Compton edge (the end of the kinematic range of the first harmonic~\cite{harvey09}), where in both linearly- and circularly-polarised backgrounds, the polarisation purity remains very high and the photons are almost entirely $E$-polarised as shown in Fig.~\ref{Fig_energy_Angle}(b). The multiple peaks in Fig.~\ref{Fig_energy_Angle}(b) correspond to different orders of harmonics.
For a circularly-polarised background, the Compton edge is at $s=2\eta/(2\eta+1+\xi^2)$ (for a linearly-polarised background $\xi^{2}\to \xi^{2}/2$). 
Furthermore, as shown in Fig.~\ref{Fig_energy_Angle}(c) and Fig.~\ref{Fig_energy_Angle}(e), the photons in the $E$-polarised states are tightly collimated with the electron's incident direction ($\vartheta_i=0$) with an angular spread of $\sim2\xi/\gamma_p$, where $\gamma_p\approx2\times 10^{5}$ is the Lorentz factor of the initial electron, in contrast to the broader angular spread of $B$-polarised photons shown in Fig.~\ref{Fig_energy_Angle}(d) and Fig.~\ref{Fig_energy_Angle}(f).
The high $E$-polarised purity can be explained with a straightforward classical multipole expansion, since the electron recoil parameter $\chi_{p} =\eta \xi\approx 0.05 \ll 1$.
When the laser intensity is in the intermediate range $\xi\not\ll 1$, $\xi \not\gg 1$, the electron produces mainly (Lorentz-boosted) dipole radiation, with some higher-order multipole contributions.
The dipole radiation is completely polarised in the parallel state, and when the electron is highly relativistic, this radiation is emitted within a narrow opening angle parallel to the electron propagation direction at the Lorentz-boosted first harmonic energy. Multipole radiation also contributes but is suppressed directly in the higher energy region [as seen in Fig.~\ref{Fig_energy_Angle}(a) and Fig.~\ref{Fig_energy_Angle}(b)] and hence there is a high purity of $E$-polarised photons at the Compton edge. A similar explanation applies to the circularly-polarised case. (To preserve azimuthal symmetry, an expansion spherical harmonics $Y^{m}_{l}$ would be dominated by terms with equal degree $l$ and order $m$, and only $Y^{0}_{0}$ (which is only present in the first harmonic) has a non-zero contribution on-axis.).

Based on the above observations we present a robust scheme to generate highly-polarised GeV $\gamma-$rays. Since $B$-polarised photons are more likely to be emitted at larger angles and lower energies, by applying an angular cut (through the placement of the detector) and an energy cut (through the use of an attenuating filter), the photons that remain are of a high $E$-polarisation purity.

\begin{figure}[t!!]
 \includegraphics[width=0.48\textwidth]{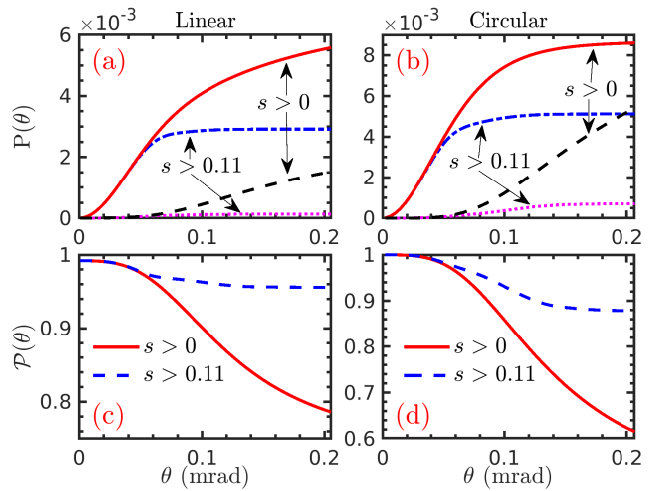}
\caption{Photon production probability $P$ and polarisation purity $\mathcal{P}$ for different detector angular acceptance values, $\theta$, for the same parameters as in Fig.~\ref{Fig_energy_Angle}. Left column: linearly polarised background. Right column: circularly polarised background. In 
(a) and (b) is the probability of a single electron scattering an $E$-polarised photon for $s>0$ (red sold line), $s>0.11$ (blue dot-dashed line) and probability for a $B$-polarised photon with $s>0$ (black dashed line), $s>0.11$ (magenta dotted line). In  (c) and (d) is the polarisation purity for $s>0$ (red sold line), $s>0.11$ (blue dashed line).}
\label{Fig_lin_Angular_Filter}
\end{figure}

From Eq.~(\ref{eqn:sfi1}), the number of photons within the detector angular acceptance $\theta$ 
can be calculated:
\begin{align}
\tsf{P}_{j}=\int^{\theta/2}_{-\theta/2}~d\theta_x \int^{\theta/2}_{-\theta/2} d\theta_y \int^{1}_{s_d} ds \frac{dr_x}{d\theta_x}\frac{dr_y}{d\theta_y}\frac{d^{3}\tsf{P}_{j}}{ds d\bm{r}^2}\,,
\label{Eq_number_angle_window}
\end{align}
where $s_d$ is a lower bound on the photon energy.
In Figs.~\ref{Fig_lin_Angular_Filter} (a) and (b), we see that the total number of detectable photons decreases with a narrower acceptance angle $\theta$. However, as the decrease of $E$-polarised photons is much slower than the decrease of $B$-polarised photons,
the polarisation purity increases sharply by narrowing the detector acceptance angle as shown in Fig.~\ref{Fig_lin_Angular_Filter} (c) and (d). For a linearly polarised background (left column), the polarisation purity of the received photons increases from $\mathcal{P}\approx 78\%$ within the acceptance angle $\theta=0.2~\textrm{mrad}$ to $90\%$ within $\theta=0.1~\textrm{mrad}$, and increases still further to the high purity $\mathcal{P}\approx 98\%$ within $\theta=0.05~\textrm{mrad}$. For the circular case (right column), the same phenomenon is presented: the polarisation purity $\mathcal{P}$ increases from $62\%$ with $\theta=0.2~\textrm{mrad}$ to $85\%$ with $\theta=0.1~\textrm{mrad}$ and to higher than $98\%$ within  $\theta=0.05~\textrm{mrad}$.
Furthermore, this high polarisation purity is carried mainly by high-energy photons because $E$-polarised photons dominate the high-energy spectrum. In Fig.~\ref{Fig_lin_Angular_Filter} we also present results for high-energy photons with an energy cutoff of $s_d=0.11$ (corresponding to $1.1~\textrm{GeV}$), where, as shown, the number  is already saturated at an acceptance $\theta\approx0.14~\textrm{mrad}$, with purity above $\mathcal{P}\approx 94\%$ ($\mathcal{P}\approx 88\%$) in a linearly-polarised (circularly-polarised) background [blue dashed line in Fig.~\ref{Fig_lin_Angular_Filter} (c) and (d)]. More than $90\%$ of these high-energy photons are collimated in a much narrower angular divergence $<0.08\,\textrm{mrad}$ with the purity above $\mathcal{P}\approx 96\%$ in each case, and more than $30\%$ of these photons are above the Compton edge in Fig.~\ref{Fig_energy_Angle} (a) and hence from the nonlinear interaction of the electron with the laser background.

In order to obtain highly brilliant $\gamma$-rays, a bunch of energetic electrons is needed to pump the photon source. Here we consider electrons with an average energy of $10~\textrm{GeV}$, attainable in upcoming LUXE \cite{Abramowicz:2019gvx} and E320 experiments \cite{slacref1} (all-optical set-ups using laser wakefield acceleration can achieve energies of order $5~\trm{GeV}$ \cite{PRL2014MultiGev,PRL2019Petawatt}). 
Because the wavelength of emitted photons is much smaller than the electron bunch length, we assume the emission is incoherent \cite{king18e}. We also assume that the electron bunch width is much smaller than the laser pulse focal width, as is planned for LUXE (strong laser focussing is not required as intermediate intensities are comfortably attainable by modern high-power lasers).
\begin{figure}[t!!]
 \includegraphics[width=0.49\textwidth]{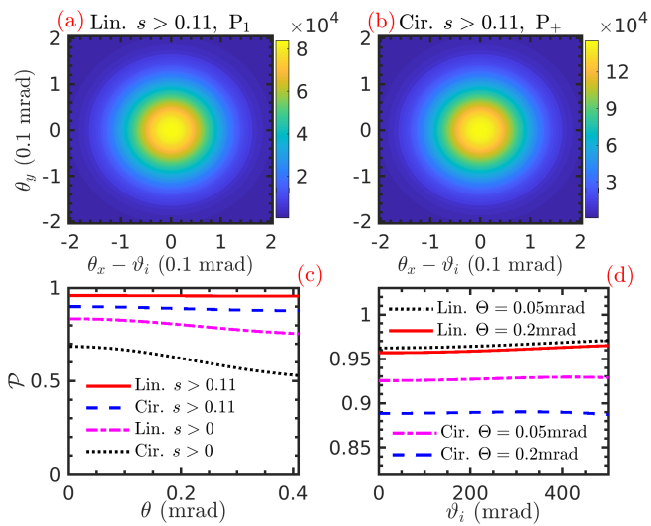}
\caption{For the same field parameters as in Fig.~\ref{Fig_energy_Angle}, with the photon detector directly along the electron propagation axis: (a) and (b) angular distribution of the high-energy ($s>0.11$ corresponding to $1.1$\,GeV) $E$-polarised photons for a linearly and circularly polarised background respectively; (c) polarisation purity $\mathcal{P}$ for varying detector acceptance $\theta$ where the pump electron beam ($\Theta=0.2~\textrm{mrad}$) collides with the laser pulse obliquely ($\vartheta_i=100~\textrm{mrad}$);
(d) polarisation purity $\mathcal{P}$ of the high-energy photons ($s>0.11$) for varying incident angle $\vartheta_{i}$ of the electron beam with the divergence $\Theta=0.2$ and $0.05~\textrm{mrad}$ respectively, where the photon detector acceptance angle is fixed equal to the beam angular divergence $\theta=\Theta$.}
\label{Fig_Bunch}
\end{figure}

In Fig.~\ref{Fig_Bunch} (a) and (b), we show the angular distribution of high-energy ($s>0.11$) $E$-polarised photons emitted by a pump electron bunch of a $6\%$ root-mean-square energy spread and an angular divergence $\Theta=0.2~\textrm{mrad}$~\cite{PRL2019Petawatt} incident in the direction $(\sin\vartheta_{i},0,-\cos\vartheta_{i})$ at an angle $\vartheta_{i}=100\,\textrm{mrad}$. (The number of electron is normalised to be unity.)
Because the bunch angular divergence is much larger than the angular spread $2\xi/\gamma_p$ induced by the background field, the photons are emitted in a much broader distribution of angles than the single-electron result in Fig.~\ref{Fig_energy_Angle}.
Therefore, rather than producing a well-defined angular harmonic structure, for a bunch of electrons with a broad spread of momenta, we see that the angular harmonic structure is smoothed out.
However, as in Fig.~\ref{Fig_energy_Angle}, we still see the dominance of the $E$-polarised photons in the high-energy region $s>0.11$. The polarisation purity is about $96\%$ ($89\%$) for the linear (circular) case shown in Fig.~\ref{Fig_Bunch} (c).
The effect of the electron bunch is to average out the peak purity along the propagation axis, resulting in a flatter distribution which is less sensitive to the precise value of the acceptance angle.
Because low-energy photons mix with higher energy photons off-axis, the maximum predicted purity is reduced, (indicated by the lines for $s>0$ in Fig.~\ref{Fig_Bunch} (c)). To reach a higher purity, an energy filter would also have to be applied. However, we emphasise that for the application of the $\gamma$-rays to measuring vacuum birefringence and generating spin-polarised particle beams, an energy cut is unnecessary as high-energy photons naturally scatter with a higher probability than low-energy photons and hence the photons that participate in the process naturally have a higher purity.

We now estimate the brilliance of the high-energy ($s>0.11$) $E$-polarised photon source. The angular divergence of the photon beam is determined by the electron bunch.
The total number of photons: $N_{\gamma}\approx 1.01\times10^{4}$ and $ 1.76\times10^{4}$, is obtained by integrating the angular distributions in Figs.~\ref{Fig_Bunch} (a), (b) respectively over the electron bunch divergence $\Theta$ and multiplying the total number of the electrons $N_e=5\times 10^{6}$ in the bunch with a density $n_e\approx1.33\times 10^{17}~\textrm{cm}^{-3}$~\cite{PRL2019Petawatt}, radius $2\lambda_l$ and length $3\lambda_l$~\cite{yanfei19}, where $\lambda_l=1\,\mu$m is the laser wavelength.
We can estimate the transverse size of the photon source to be comparable to the size of the electron bunch and its duration equal to the half of the sum of the bunch length $t_e=3\lambda_l/c=10~\textrm{fs}$ and the laser duration: $t_l=2\sigma \lambda_l=26.7~\textrm{fs}$.
The photon number in a $0.1\%$ bandwidth (BW) is obtained by averaging the total number of the photons.
We then acquire a prediction for the brilliance for linearly and circularly polarised backgrounds of, respectively, $1.1\times10^{21}$ and $1.9\times10^{21}$ $\textrm{photons}/(\textrm{s}~\textrm{mm}^2~\textrm{mrad}^2~0.1\%~\textrm{BW})$, which is more than $3$ orders of magnitude brighter than recently suggested high energy polarised photon sources~\cite{yanfei19}.
This high brilliance and improvement of the polarisation purity on previous suggestions can be explained as follows.
First, our scheme exploits photons produced around the edge of the first harmonic, which carries a higher purity of $E$-polarisation than elsewhere in the spectrum, and it is well known that the harmonics (particularly at the Compton edge) cannot be described by the locally-constant-field approximation that is central to numerical simulations employed in other analyses \cite{ritus85,king15d,dipiazza18,king19a}. Second, at this lower intensity the angular divergence is extremely narrow and the field-induced angular spread of the electrons is negligible.

The robustness of our scheme is tested by varying the incident angle $\vartheta_i$ of the electron beam with a detector directly downstream, set to a fixed acceptance angle $\theta=\Theta$. We see in Fig.~\ref{Fig_Bunch} (d) that the high polarisation purity of the high-energy photons is maintained for a broad range of incident angles. As the bunch incident angle $\vartheta_i$ is increased from $0$ to $200~\textrm{mrad}$, the brilliance decreases less than $5\%$ (see Supplemental Material in the appendix for details on the calculation for an electron bunch.). One reason for such a small decrease is the relative insensitivity of the energy parameter $\eta$ to small change in the collision angle from head-on alignment [$\eta \propto \vkap\cdot p\propto 1+\cos(\vartheta_i)$]. (Further increase of the incident angle to $0.5\,$rad results in a faster decay of the brilliance (see Supplementary Material in the appendix).)

One way to further improve the brilliance and polarisation purity is to reduce the angular divergence of the electron bunch. If the angular divergence of the electron beam is reduced to the same level of the field-induced angular spread: $\Theta=2\xi/\gamma_p\approx0.05~\textrm{mrad}$, the bunch spectra will revert to the single-particle results in Figs.~\ref{Fig_energy_Angle} and~\ref{Fig_lin_Angular_Filter}, and the harmonic structure realised (see Supplementary Material in the appendix). In this way, the brilliance of the high-energy photons $s>0.11$ can be improved to above $10^{22}~\textrm{photons}/(\textrm{s}~\textrm{mm}^2~\textrm{mrad}^2~0.1\%~\textrm{BW})$, and the corresponding polarisation purity can be improved to about $97\%$ ($93\%$) for a linearly (circularly) polarised background as shown in Fig.~\ref{Fig_Bunch} (d).
Furthermore, the photons with different polarisations are angularly separated, and thus a simple angular selection is sufficient to filter pure, highly polarised GeV $\gamma$-rays.

In conclusion, we detailed a robust scheme to generate highly polarised GeV $\gamma$-rays with ultrahigh brilliance up to $10^{21}~\textrm{photons}/(\textrm{s}~\textrm{mm}^2~\textrm{mrad}^2~0.1\%~\textrm{BW})$. Our scheme exploits the fact that, starting around the Compton edge, photons are mainly scattered along the electron's propagation axis in one polarisation state. To maximise this effect, a laser pulse with an intermediate intensity should be collided almost head-on with high-energy electrons ($\sim 10\,\trm{GeV}$). (At higher intensities, the photon angular spread becomes larger, and at lower intensities, the probability of scattering is smaller.) A direct calculation from QED is required for these parameters, as they are outside the region of applicability of numerical simulations based on the locally constant field approximation. The brilliance and polarisation purity of the photon source can be further improved by employing electron beams of higher energy and smaller divergence angle and by increasing the spatio-temporal overlap with the colliding laser pulse.

\begin{acknowledgments}
S. T. and B. K. thank A. Ilderton for a careful reading of the manuscript. They are supported by the UK Engineering and Physical Sciences Research Council, Grant No. EP/S010319/1. H.H. acknowledges the support by the National Natural Science Foundation of China under Grant No. 11774415.
\end{acknowledgments}

\appendix

\section*{Supplementary Material}

In this Supplementary Material (SM), we include some miscellaneous extra material, which is not central to the description of our results in the main text, but can provide some extra information on the nature of the calculations performed and the scope of the results. The numbers of the equations and figures in this material contain the prefix ``SM'' to distinguish the usual numbers of the equations and figures in the main text.

\subsection*{Robustness of scheme to collision angle}
In the main text, we show in Fig.~3 the angular distribution of high-energy ($s > 0.11$ corresponding to $>1.1\,\trm{GeV}$) $E$-polarised photons emitted by a pump electron bunch with an angular divergence \mbox{$\Theta=0.2~\textrm{mrad}$}. Because the bunch angular divergence is several times broader than the field-induced angular spread, \mbox{$2\xi/\gamma_p$}, the harmonic structure in the polarised-photon angular distribution is smoothed out. 

Integrating the angular distributions over a particular detector acceptance angle $\theta$ and multiplying the number of the electrons in the bunch, we can obtain the total number of photons accepted by the detector. In the main text, as an example we present a calculation of the photon source brilliance from an electron bunch incident at an angle $\vartheta_{i}=100\,$mrad with divergence $\Theta=0.2\,$mrad.

In Fig.~\ref{Fig_Brilliance}, we present the brilliance obtained for the high-energy $E$-polarised photon source for a broad range of bunch incident angles and different bunch angular divergence. As shown, the brilliance decreases less than $5\%$ as the bunch incident angle $\vartheta_i$ is increased from $0$ to $200\,\textrm{mrad}$, and further increase of the incident angle to $500\,\textrm{mrad}$ would result in a faster decay of the brilliance. 

\subsection*{Increased Brilliance of Highly Polarised Source}
\begin{figure}[t!!]
 \center{\includegraphics[width=0.44\textwidth]{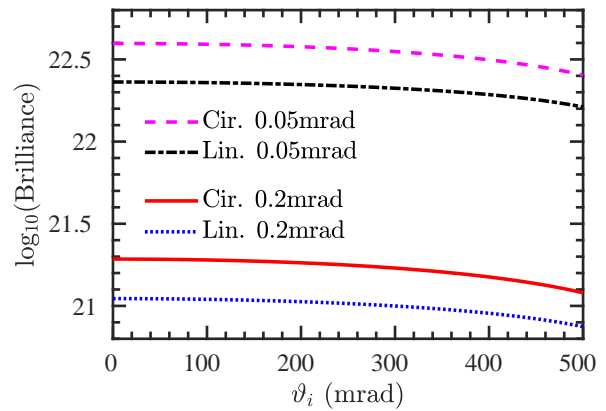}}
\caption{Brilliance of the $E$-polarised high-energy photons ($s>0.11$ corresponding to $>1.1~\textrm{GeV}$) for different incident angle $\vartheta_i$ and electron beam angular divergence $\Theta=0.05~\textrm{mrad}$ and $0.2~\textrm{mrad}$. The acceptance angle of the detector is set to $\theta=\Theta$.
The field parameters are the same as in the main text (for a laser pulse with $\xi=0.5$, $\omega_l=1.24~\textrm{eV}$ and duration $26.7\,\trm{fs}$ ($8$ cycles)). The electron bunch has a mean energy of  $10\,$GeV and a root-mean-square energy spread of $6\%$. The photon detector is set downstream along the bunch propagation direction.}
\label{Fig_Brilliance}
\end{figure}

One way to improve the brilliance of the polarised source, would be to reduce the electron bunch's angular divergence. In Fig.~\ref{Fig_Brilliance}, we show example results for an angular divergence of $\Theta=0.05\,\textrm{mrad}$.  It follows that the angular divergence of the photon beam is determined by the properties of the electron bunch and we find that the brilliance of the photon source is inversely proportional to the electron bunch divergence, $\propto\Theta^{-2}$.

\subsection*{Calculation of the angular spectrum}
As mentioned in the main text, to obtain a brilliant photon source, a high-energy electron bunch is required. The calculation we performed for the electron bunch, assumed a normalised momentum distribution of the form:
\bea
&\rho(\bm{p})=\frac{1}{\sqrt{2\pi^3}\sigma_{\parallel}\sigma^2_{\perp} m^3}\exp\left[-\frac{(\bm{p}\cdot\bm{n}-\tilde{p})^2}{2\sigma^2_{\parallel}m^2}-\frac{\left|\bm{p}-\bm{n}(\bm{p}\cdot\bm{n})\right|^2}{\sigma^2_{\perp} m^2}\right]\,,\nonumber
\eea
with modulus average momentum $\tilde{p}$ and rms momentum spread in the longitudinal, $\sigma_{\parallel} m$, and transverse, $\sigma_{\perp} m $ directions, where $\bm{n}$ is the incident direction of the electron bunch.

The differential probability of emitting a photon in the polarisation state $\eps_{j}$ with momentum $k$ is straightforwardly acquired from Eq.~(1) of the main text:
\begin{align}
\frac{d^3\tsf{P}_{j}}{ds\,d^2\bm{r}} &= \frac{\alpha}{(2\pi\eta)^{2}}\int^{\infty}_{s} \frac{d\lambda}{\lambda^2}\int d^2 p^{\LCperp} p^{0}\rho(\bm{p}) \nonumber \\
& \times \frac{s}{t}\int d\phi\,d\phi'~\tsf{T}_{j}~\e^{i\int^{\phi}_{\phi'}\frac{k\cdot \pi_{p}(\vphi)}{m^2t\eta}d\vphi}\,,\label{eqn:sfi1}
\end{align}
where $\lambda=\vkap\cdot p/m^2\eta$ is the electron light-front momentum normalised by the interaction energy parameter $\eta$ used in the manuscript, $s=\vkap\cdot k/m^2\eta$, $t=\lambda-s$. The expression of $\tsf{T}_{j}$ ($j=1,2$ for the linear case, and $j=+,-$ for the circular case) is same as in the manuscript except the dependency on $s$ and $t$ is replaced with $s/\lambda$ and $t/\lambda$, respectively.

From Eq.~(\ref{eqn:sfi1}), the angular distribution of the polarised photon can be acquired: 
\begin{align}
\frac{d^2\tsf{P}_{j}}{dr_x\,dr_y} =\frac{\alpha}{(2\pi\eta)^{2}} \int^{\infty}_{s_d} \frac{d\lambda}{\lambda^2}\int d^2 p^{\LCperp}\, p^{0}\rho(\bm{p})\,g\big(\lambda, \bm{r}-\frac{p^{\LCperp}}{m\lambda}\big)
\label{eqn:sfi2}
\end{align}
in which we use the shorthand
\begin{align}
g\left(\lambda, \bm{r}-\frac{p^{\LCperp}}{m\lambda}\right)=\int^{\lambda}_{s_d}ds\frac{s}{t}\int d\phi\,d\phi'~\tsf{T}_{j}~\e^{i\int^{\phi}_{\phi'}\frac{k\cdot \pi_{p}(\vphi)}{m^2 t \eta}d\vphi}\,.
\end{align}

\begin{figure}[t!!]
 \center{\includegraphics[width=0.48\textwidth]{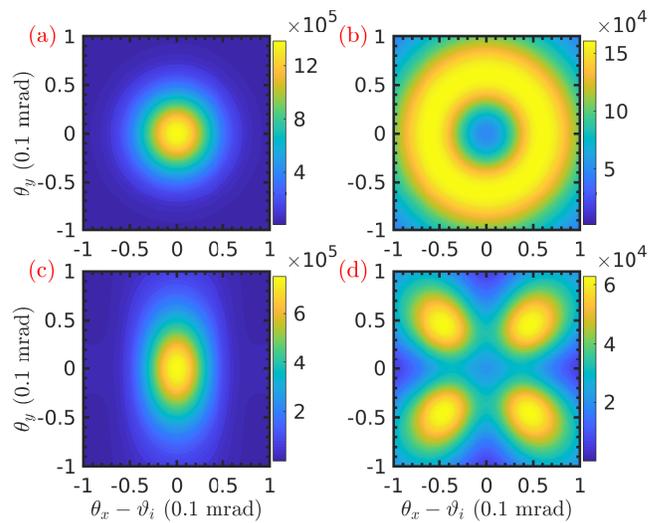}}
\caption{Angular distribution $d^2\tsf{P}/d\theta_xd\theta_y$ of the polarised photons ($s_d=0$) in a circularly-polarised background (upper panels) and a linearly-polarised background (bottom panels). Left column: (a) and (c) angular distribution of $E$-polarised photons. Right column: (b) and (d) angular distribution of $B$-polarised photons. The field parameters are same as in the main text. The parameters of electron bunch are listed: angular divergence $\Theta=0.05\,$mrad, incident angle $\vartheta_i=100\,$mrad, average energy $10$~GeV and root-mean-square energy spread $6\%$, and. The photon detector is set downstream along the bunch propagation direction.}
\label{Fig_bunch_angle_dist_005}
\end{figure}
To view the angular harmonic structure shown in Fig.~1 from the single-particle results, the bunch angular divergence has to be reduced to the same level of the angular spread induced by field, thus to be $\Theta \approx 2\xi/\gamma_p$. Fig.~\ref{Fig_bunch_angle_dist_005} shows the angular distribution of polarised photons emitted by a pump electron bunch of average energy $\tilde{p}=\gamma_p m$ ($10\,$GeV),
$\sigma_{\parallel}=0.03\gamma_p$ and $\sigma_{\perp}=2.5\times 10^{-5} \gamma_p$ corresponding to a $6\%$ root-mean-square energy spread and an angular divergence $\Theta=0.05\,\textrm{mrad}$, incident in the direction $\bm{n}=[\sin\vartheta_{i},0,-\cos\vartheta_{i}]$ at an angle $\vartheta_{i}=100\,\textrm{mrad}$. As shown in the figure, well-defined harmonic structures in the angular distribution can be clearly observed.

\bibliographystyle{apsrev}

 \providecommand{\noopsort}[1]{}

\end{document}